\documentclass[useAMS,usenatbib]{mn2e}
\usepackage{graphicx}
\usepackage{times}
\newcommand{\hompc}{\,h\,{\rm Mpc}^{-1}}
\newcommand{\mpcoh}{\,h^{-1}\,{\rm Mpc}}
\newcommand{\rev}[1]{{#1}}  

\begin{document}

\title[Including covariance matrix errors in BOSS] 
{The Clustering of Galaxies in the SDSS-III Baryon Oscillation Spectroscopic Survey: Including covariance matrix errors}

\author[Percival et al.]{
\parbox{\textwidth}{
Will J. Percival$^{1}$\thanks{E-mail: will.percival@port.ac.uk}, 
Ashley J. Ross$^{1}$,
Ariel G. S\'anchez$^{2}$,
Lado Samushia$^{1,3}$,
Angela Burden$^{1}$,
Robert Crittenden$^{1}$,
Antonio J. Cuesta$^{4}$,
Mariana Vargas Magana$^{5}$,
Marc Manera$^{1}$,
Florian Beutler$^{6}$,
Chia-Hsun Chuang$^{7}$,
Daniel J. Eisenstein$^{8}$,
Shirley Ho$^{5}$,
Cameron K. McBride$^{8}$,
Francesco Montesano$^{2}$,
Nikhil Padmanabhan$^{2}$,
Beth Reid$^{6}$,
Shun Saito$^{6,9}$,
Donald P. Schneider$^{10,11}$,
Hee-Jong Seo$^{6}$,
Rita Tojeiro$^{1}$,
Benjamin A. Weaver$^{12}$,
}\\
\vspace*{4pt} \\
$^{1}$ Institute of Cosmology and Gravitation, University of
          Portsmouth, Dennis Sciama Building, Portsmouth, P01 3FX, UK  \\
$^{2}$ Max-Planck-Institut fur Extraterrestrische Physik,
          Giessenbachstraße, D-85748 Garching, Germany \\
$^{3}$ National Abastumani Astrophysical Observatory, Ilia State
          University, 2A Kazbegi Ave., GE-1060 Tbilisi, Georgia  \\
$^{4}$ Department of Physics, Yale University, 260 Whitney Ave, New
          Heaven, CT 06520, USA  \\
$^{5}$ Bruce and Astrid McWilliams Center for Cosmology, Department of
          Physics, Carnegie Mellon University, 5000 Forbes Ave,
          Pittsburgh, PA 15213, USA  \\
$^{6}$ Lawrence Berkeley National Laboratory, One Cyclotron Road,
          Berkeley, CA 94720, USA \\
$^{7}$ Instituto de Fısica Teorica, (UAM/CSIC), Universidad Autonoma
         de Madrid, Cantoblanco, E-28049 Madrid, Spain \\
$^{8}$ Harvard-Smithsonian Center for Astrophysics, Harvard
          University, 60 Garden St., Cambridge, MA 02138, USA \\
$^{9}$ Kavli Institute for the Physics and Mathematics of the Universe
          (WPI), Todai Institues for Advanced Study, The University of Tokyo,
          Chiba 277-8582, Japan \\
$^{10}$ Department of Astronomy and Astrophysics, The Pennsylvania
          State University, University Park, PA 16802, USA\\
$^{11}$ Institute for Gravitation and the Cosmos, The Pennsylvania
          State University, University Park, PA 16802, USA\\
$^{12}$ Center for Cosmology and Particle Physics, New York
          University, New York, NY 10003 USA}

\date{\today} 
\pagerange{\pageref{firstpage}--\pageref{lastpage}} \pubyear{2013}
\maketitle
\label{firstpage}

\begin{abstract}
  We present improved methodology for including covariance matrices in
  the error budget of Baryon Oscillation Spectroscopic Survey (BOSS)
  galaxy clustering measurements, \rev{revisiting Data Release 9 (DR9)
  analyses, and describing a method that is used in DR10/11 analyses
  presented in companion papers.} The precise analysis method adopted
  is becoming increasingly important, due to the precision that BOSS
  can now reach: even using as many as 600 mock catalogues to estimate
  covariance of 2-point clustering measurements can still lead to an
  increase in the errors of $\sim$20\%, depending on how the
  cosmological parameters of interest are measured. In this paper we
  extend previous work on this contribution to the error budget,
  deriving formulae for errors measured by integrating over the
  likelihood, and to the distribution of recovered best-fit parameters
  fitting the simulations also used to estimate the covariance
  matrix. Both are situations that previous analyses of BOSS have
  considered. We apply the formulae derived to Baryon Acoustic
  Oscillation (BAO) and Redshift-Space Distortion (RSD) measurements
  from BOSS\rev{ in our companion papers. To further aid these
    analyses, we consider} the optimum number of bins to use for 2-point
  measurements using the monopole power spectrum or correlation
  function for BAO, and the monopole and quadrupole moments of the
  correlation function for anisotropic-BAO and RSD measurements.
\end{abstract}


\section{Introduction}  \label{sec:intro}

With the increasing precision enabled by modern cosmological
observations (e.g. \citealt{anderson12,planck13}), there is increasing
interest in making their statistical analysis as rigorous as the
measurements themselves. In this brief paper we review the propagation
of errors in the covariance matrix to the parameter errors, extending
recent work (\citealt{taylor12,dodelson13}) to cover errors estimated
by marginalising over the likelihood recovered for each mock, and
errors measured from the distribution of mocks that are also used to
estimate the covariance matrix. These situations arose in our recent
analysis measuring the Baryon Acoustic Oscillation (BAO) postion in
the Baryon Oscillation Spectroscopic Survey (BOSS;
\citealt{eisenstein11,dawson13}) Data Release 9 (DR9; \citealt{ahn12})
galaxy samples \citep{anderson12}, and in related analyses. 

Many cosmological observations are well described as being drawn from
a multi-variate Gaussian distribution with inverse covariance matrix
$\Psi^t$, where the superscript $t$ denotes the true matrix, so that
parameter inferences (such as finding the BAO position) can be based
on a likelihood
\begin{equation}  \label{eq:like}
  {\cal L}({\bf x}|{\bf p},\Psi^t)
    =\frac{|\Psi^t|}{\sqrt{2\pi}}\exp\left[-\frac{1}{2}\chi^2({\bf x},{\bf p},\Psi^t)\right],
\end{equation}
where
\begin{equation} \label{eq:chisq}
  \chi^2({\bf x},{\bf p},\Psi^t) \equiv \sum_{ij} 
    \left[x_i^d-x_i({\bf p})\right]\Psi^t_{ij}\left[x_j^d-x_j({\bf p})\right].
\end{equation} 
In the example of BAO fitting, the data ${\bf x}^d$, and model for the
data ${\bf x}({\bf p})$, would be power spectra or correlation
functions, with the parameter ${\bf p}$ being the BAO position.

In many experiments, it is common to use mock, or simulated, data to
estimate the inverse covariance matrix $\Psi^t$. Suppose we have $n_b$
data measurements such as power spectrum band-powers, and wish to
estimate the covariance matrix using $n_s$ simulations.  Assuming that
the mock data can be written as $x^s_i$, with $1\le i\le n_b$ and
$1\le s\le n_s$, the mean of each value over all simulations is
\begin{equation}
  \mu_i=\frac{1}{n_s}\sum_s x^s_i,
\end{equation}
and an unbiased estimate of the true covariance matrix $C^t$ from these data is
\begin{equation}  \label{eq:cov_estimate}
  C_{ij} = \frac{1}{n_s-1}\sum_s (x^s_i-\mu_i)(x^s_j-\mu_j).
\end{equation}
The distribution of matrices recovered from multiple, independent sets
of simulations follows the statistics of a Wishart distribution, and
its inverse $\Psi$, from an inverse-Wishart distribution with true
inverse covariance matrix $\Psi^t$ (e.g. \citealt{press05}).

Because we do not know $\Psi^t$, we cannot use Eq.~(\ref{eq:like})
directly, but should instead make parameter inferences using a joint
likelihood
\begin{equation}  \label{eq:like_full}
  {\cal L}({\bf x},\Psi|{\bf p},\Psi^t) = {\cal L}({\bf x}|{\bf p},\Psi) {\cal L}(\Psi|\Psi^t),
\end{equation}
where ${\cal L}(\Psi|\Psi^t)$ is given by an inverse Wishart
distribution, while ${\cal L}({\bf x}|{\bf p},\Psi)$ is the standard
distribution given in Eq.~(\ref{eq:like}), after replacing the true
inverse covariance matrix with the estimate. We can subsequently
marginalise over $\Psi^t$ to obtain ${\cal L}({\bf x},\Psi|{\bf
  p})$, which can be used to derive parameter measurements.

The marginalisation over all elements in $\Psi^t$ is computationally
challenging; this limitation has led to an approximate approach, where
the estimate of $\Psi^t$ is used instead of the true inverse
covariance matrix in Eq.~(\ref{eq:like}), and the method and results
from this approach are corrected. Marginalising over the distribution
of measured covariance matrices in Eq.~(\ref{eq:like_full}) leads to
two important corrections to this simplified approach:
\begin{enumerate}
\item The inverse Wishart distribution has a form such that $C^{-1}$,
  with $C$ determined as in Eq.~(\ref{eq:cov_estimate}), is a biased
  estimate of the inverse covariance matrix.
\item The marginalisation over possible true inverse covariance
  matrices increases the width of the error on any measured parameter
  from that recovered from ${\cal L}({\bf x}|{\bf p},\Psi)$.
\end{enumerate}

The first effect can be corrected by using an unbiased estimate of the
inverse covariance matrix in the likelihood calculation
\begin{equation}  \label{eq:Psi}
  \Psi = (1-D) C^{-1}, \,\,\,\,\, D=\frac{n_b+1}{n_s-1} 
\end{equation}
where the factor $D$ accounts for the skewed nature of the inverse
Wishart distribution (for the first cosmological application of this,
see \citealt{hartlap07}). 

Changing the covariance matrix in this manner does not correct for
errors in the covariance matrix, which propagate through to errors on
estimated parameters, so the second effect is still apparent. Suppose
that the inverse covariance matrix estimate has an error $\Delta\Psi$
compared with the true matrix $\Psi^t$, with $\Psi =
\Psi^t+\Delta\Psi$. For simulations drawn from a multi-variate
Gaussian, these errors can be calculated \citep{taylor12},
\begin{equation}  \label{eq:DPsi}
  \langle \Delta\Psi_{ij}\Delta\Psi_{i'j'}\rangle
    = A\Psi_{ij}\Psi_{i'j'}+B(\Psi_{ii'}\Psi_{jj'}+\Psi_{ij'}\Psi_{ji'}),
\end{equation}
where
\begin{eqnarray}  
  A&=&\frac{2}{(n_s-n_b-1)(n_s-n_b-4)},\nonumber\\ 
  B&=&\frac{(n_s-n_b-2)}{(n_s-n_b-1)(n_s-n_b-4)}. \label{eq:AB}
\end{eqnarray}

In the following three sections, we consider how to use these error
estimates to correct various parameter error calculations in order to
fully account for the errors in the covariance matrix. In
Section~\ref{sec:full_err}, we first follow the derivation of
\citet{dodelson13}, calculating the true error for measurements made
from data that are independent from that used to estimate the
covariance matrix. In Section~\ref{sec:like}, we consider how the
covariance-matrix errors propagate through to an estimate of the
confidence interval derived from an individual likelihoods, and how
measurements made from this approach must be corrected to give the
true error. Section~\ref{sec:dist_same_data} considers the
distribution of values recovered when fitting the same simulated data
used to estimate the covariance: this exercise serves as a test of the
method, allowing the full set of simulations to be used to both create
and test the covariance matrix estimate.  For consistency and brevity
in these sections, we follow the notation of \citet{dodelson13} as
closely as possible. The derived formulae are tested using Monte-Carlo
simulations in Section~\ref{sec:mc_tests}.

While following the propagation of errors in the covariance matrix
through to parameter errors ensures that the estimated parameter
errors are unbiased, this calculation does not mean that the corrected
${\cal L}({\bf x}|{\bf p},\Psi)$ provides a maximum likelihood
estimator for ${\bf p}$. Instead, the corrected parameter errors
depend on the number of bins used when modelling the data, which can
be considered as part of the methodology: smaller values of $n_b$ give
rise to less noisy estimates of the covariance matrix elements, while
we cannot determine the elements of larger covariance matrices with
the same precision. Using larger covariance matrices leads to
increasingly large deviations in the accuracy of the parameter
measurements compared with those that would have been made using the
true likelihood. In Section~\ref{sec:boss} we provide a practical
demonstration of the corrections, calculating the optimal number of
bins to use when performing cosmological analyses of the latest BOSS
galaxy clustering data.

\section{The combined error}  \label{sec:full_err}

Suppose that we have estimated the covariance matrix using a sample of
simulations, and wish to know the full error that we should expect on
a measurement made using this covariance matrix and the standard
Gaussian likelihood, or equivalently the expected distribution of
best-fit parameter values that would be recovered from an independent
set of simulations. This calculation was performed by
\citet{dodelson13}, and corresponds to determining the combined error
on a measurement including both the data and covariance matrix errors.

We assume that the likelihood is calculated using the inverse
covariance matrix estimate of Eq.~(\ref{eq:Psi}). Following
equation~24 of \citet{dodelson13}, and using the standard summation
convention, we can write the estimator for parameter $p_\alpha$ as
\begin{equation}  \label{eq:p_alpha}
  \hat{p}_\alpha=[F+\Delta F]^{-1}_{\alpha\alpha'}
    \frac{\partial x_i}{\partial p_{\alpha'}}\Psi_{ij}(x_j^d-x_j^t),
\end{equation}
where $F$ is the true Fisher matrix linearly relating the fitted
parameter $p$ to the measurements around the true likelihood peak
\begin{equation}  \label{eq:F}
  F_{\alpha\beta} \simeq\sum_{ij}
    \frac{\partial x_i}{\partial p_{\alpha}} \Psi^t_{ij}
    \frac{\partial x_j}{\partial p_{\beta}},
\end{equation}
and similarly for $\Delta F$ as a function of $\Delta\Psi$. Also
following \citet{dodelson13}, and without loss of generality, we
assume that the true values of the parameters $p_{\alpha}=0$.

\citet{dodelson13} expanded Eq.~(\ref{eq:p_alpha}) to find the second
 order (s.o.) contribution to the expected distribution of recovered
 values.
\begin{equation}
  \left.\langle p_\alpha p_\beta
    \rangle\right|_{s.o.}=B (n_b-n_p) F_{\alpha\beta}^{-1},
\end{equation}
where $B$ was given in Eq.~(\ref{eq:AB}), and $n_p$ is the number of
parameters measured. Thus, the corrected variance is
\begin{equation}  \label{eq:var_idist}
  V_{\alpha\beta}=\left[1+B(n_b-n_p)\right]F^{-1}_{\alpha\beta}.
\end{equation}
This result, which was a key conclusion of \citet{dodelson13},
describes the additional contribution to the data error from a
covariance matrix calculated from simulations. It matches the
distribution of best-fit parameter measurements made from a set of
simulations that is independent of those used to estimate the
covariance matrix. However, this correction cannot be directly applied
to an error derived from the likelihood derived from a particular
mocks (as made in \citealt{anderson12}, for example), as is
demonstrated in the next section.

\section{Errors from the likelihood}  \label{sec:like}

In order to propagate the uncertainty in the covariance matrix through
to errors estimated from the recovered likelihood for a particular
fit, we first review how these errors are usually calculated. The
best-fit measurement can be made by integrating over the likelihood
\begin{equation}
  \hat{p}_\alpha=\int \frac{p_\alpha}{\sqrt{2\pi|\Psi^{-1}|}} 
    \exp{-\frac{1}{2}\chi^2({\bf x},{\bf p},\Psi)}\,dp,
\end{equation}
with $\chi^2$ defined as in Eq.~(\ref{eq:chisq}).  In the
multi-variate Gaussian approximation around the best-fit solution,
this expression reduces to Eq.~(\ref{eq:p_alpha}).

The (squared) error on the measurement can also be estimated by
integrating over the likelihood,
\begin{equation}  \label{eq:sig_from_like}
  \hat{\sigma}^2_{\alpha\beta}=
  \int\frac{(p_\alpha-\hat{p}_\alpha) (p_\beta-\hat{p}_\beta)}{\sqrt{2\pi|\Psi^{-1}|}}
  \exp{-\frac{1}{2}\chi^2({\bf x},{\bf p},\Psi)}\,dp.
\end{equation}
If $\Psi$ were known perfectly (i.e., we replace $\Psi$ with $\Psi^t$),
Eq.~(\ref{eq:sig_from_like}) would recover a parameter variance of
$[F]^{-1}_{\alpha\beta}$, from the definition of the Fisher matrix. The
error in $\Psi$ instead leads to a revised variance estimate
\begin{equation} \label{eq:var_step1}
  \hat{\sigma}^2_{\alpha\beta}=[F+\Delta F]^{-1}_{\alpha\beta}.
\end{equation}
A Taylor series expansion then gives
\begin{equation} \label{eq:var_step2}
  \hat{\sigma}^2_{\alpha\beta}=F^{-1}_{\alpha\beta}
    +(F^{-1}\Delta F\,F ^{-1}\Delta F\,F^{-1})_{\alpha\beta},
\end{equation}
ignoring first order terms that will lead to zero expectation. Using
the analogue of Eq.~(\ref{eq:F}) for $\Delta F$ as a function of
$\Delta\Psi$, and substituting in Eq.~(\ref{eq:DPsi}), we see that the
error from the covariance matrix estimation increases the recovered
variance to yield
\begin{equation}  \label{eq:var_like}
  \hat{\sigma}^2_{\alpha\beta}=\left[1+A+B(n_p+1)\right]F^{-1}_{\alpha\beta}.
\end{equation}
Thus, the error in the covariance matrix has a biased effect on errors
derived from the likelihood from any particular fit: on average they
are larger than the errors would have been if we knew the true inverse
covariance matrix.  Unfortunately, the increase in size does not match
the increase required to correct the distribution of best-fit values
recovered from independent data as derived by \citet{dodelson13}, and
presented in the previous section. To obtain an unbiased estimate of
the full variance on parameter $p_\alpha$, given a measurement of the
error made using the standard method of integrating over the
likelihood, we therefore must apply a factor of
\begin{equation}  \label{eq:m_1}
  m_1 = \frac{V_{\alpha\beta}}{\hat{\sigma}^2_{\alpha\beta}}
  = \frac{1+B(n_b-n_p)}{1+A+B(n_p+1)},
\end{equation}
to the measured parameter covariance, and the square root of this
expression to the measured standard deviation.

Because the correction to the measured parameter covariance is
independent of the value of the parameters around which the variance
is measured, this correction should be applied even if we wish to
estimate errors from the recovered likelihood, calculated by fitting
to the same data used to estimate the covariance matrix.

\section{Distribution of same data}  \label{sec:dist_same_data}

In general, one wants to construct the best covariance matrix
possible, in order to minimise the additional error. Thus, if this
matrix is to be based on simulations, it is strongly desirable to use
all available simulations. A classical approach is to apply any data
analysis pipeline to mock data in order to test for any problems. If
all mocks have already been used to estimate the covariance matrix,
however, we should not expect to recover a distribution of best-fit
solutions that matches the equations derived in
Section~\ref{sec:full_err}.

Consequently, it is worth examining how the expected error changes
when we analyse the distribution of best-fit values recovered from the
same data set used to estimate the covariance matrix. In this case, we
can write
\begin{equation}
  \langle (x_i^d-x_i^t) (x_j^d-x_j^t) \rangle
    = (1-D)(\Psi^{-1})_{ij},
\end{equation}
from Eqns.~(\ref{eq:cov_estimate}) \&~(\ref{eq:Psi}). Substituting
this equation into an expansion of $\langle p_\alpha p_\beta \rangle$,
with $p_\alpha$ as in Eq.~(\ref{eq:p_alpha}), we find
\begin{equation}
  \langle p_\alpha p_\beta
    \rangle= (1-D)[F+\Delta F]^{-1}_{\alpha\beta}.
\end{equation}
Using the same approach that led from Eq.~(\ref{eq:var_step1}) to
Eq.~(\ref{eq:var_like}), yields
\begin{equation}
  \langle p_\alpha p_\beta \rangle = 
    \left[(1-D)(1+A+B(n_p+1))\right]F^{-1}_{\alpha\beta}.
\end{equation}
Therefore, the distribution of best-fit parameter values recovered
from data that was also used to estimate the covariance matrix is
biased in a different way to that of an independent set of data, and
from the covariance estimate made from the measured
likelihood. However, we can still use the recovered distribution to
test the methodology provided we include the revised bias when
analysing the result. Here we need a corrective factor
\begin{equation}  \label{eq:m_2}
    m_2 = \frac{V_{\alpha\beta}}{\langle p_\alpha p_\beta \rangle} = (1-D)^{-1}m_1,
\end{equation}
with $m_1$ defined as in Eq.~(\ref{eq:m_1}).

\section{Testing using Monte-Carlo simulations}  \label{sec:mc_tests}

\begin{figure}
  \centering
  \resizebox{0.78\columnwidth}{!}{\includegraphics{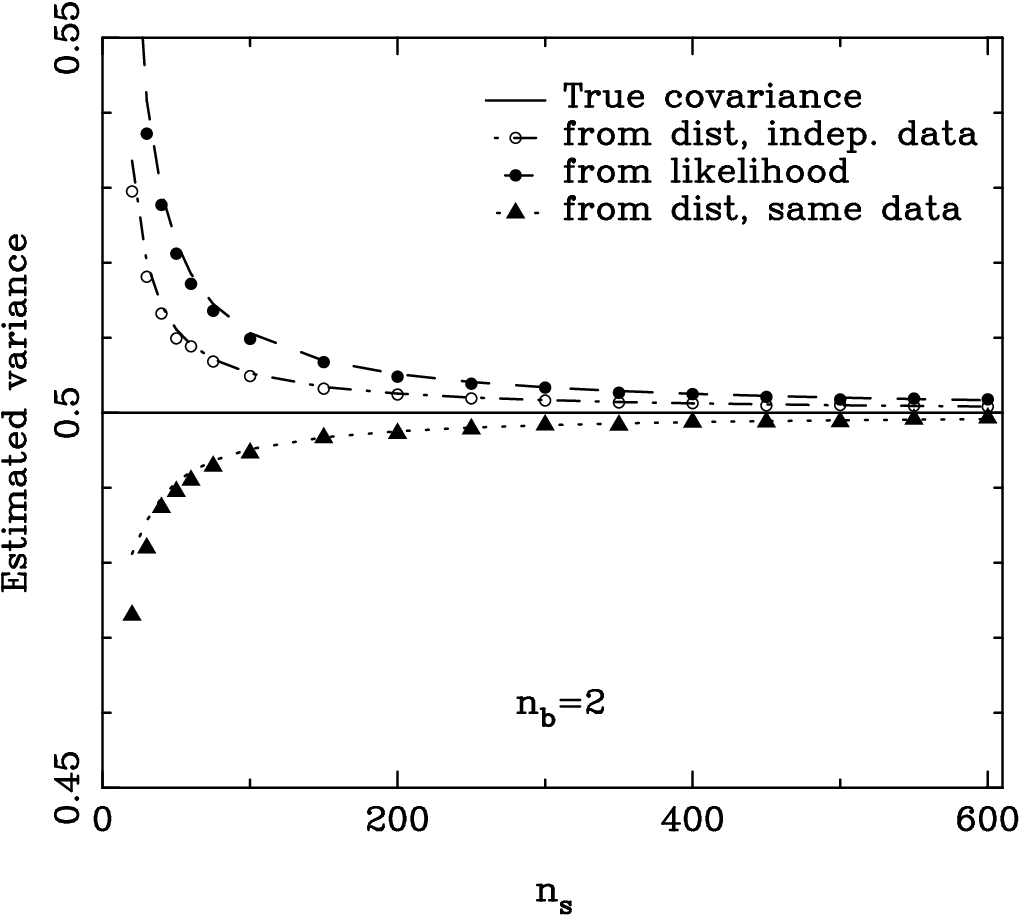}}
  \resizebox{0.78\columnwidth}{!}{\includegraphics{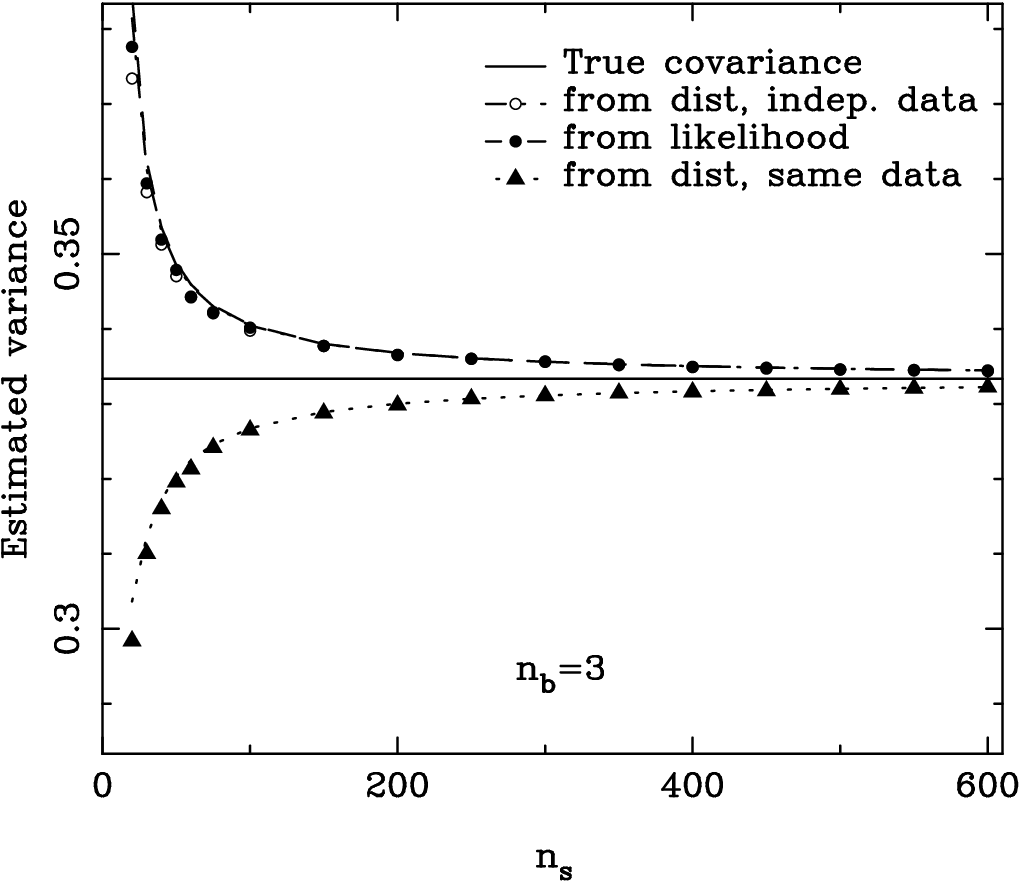}}
  \resizebox{0.78\columnwidth}{!}{\includegraphics{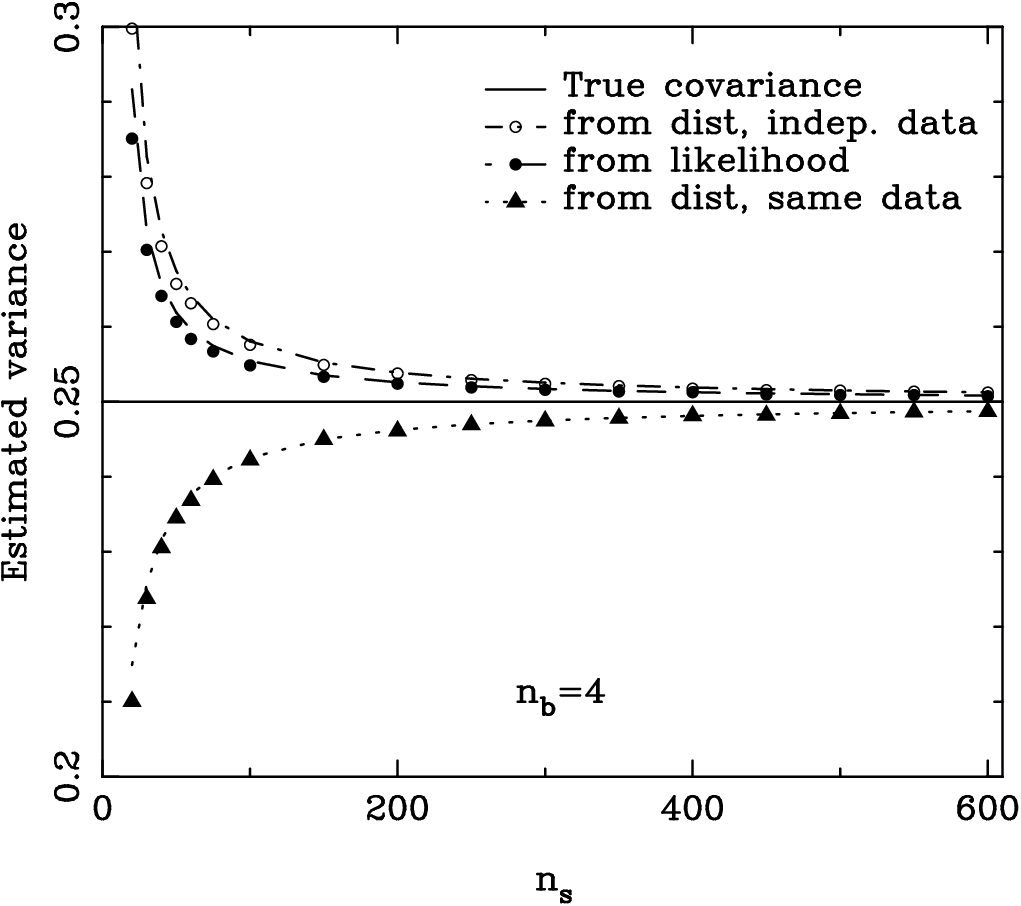}}
  \caption{Estimated variance for the mean of $n_b$ independent
    standard Gaussian random variables. The symbols show the estimated
    variance, averaged over $10^5$ runs, each using $n_s$ data vectors
    to calculate the covariance matrix. Solid circles show the average
    variance, calculated from the $n_s$ likelihood distributions
    derived fitting to $n_s$ independent data vectors (see
    Section~\ref{sec:like}), open circles from the distribution of
    best-fit solutions recovered from these data (see
    Section~\ref{sec:full_err}), and the solid triangles from the
    distribution of best-fit solutions when the same data used to
    estimate the covariance matrix is fitted (see
    Section~\ref{sec:dist_same_data}). No corrections were applied to
    these estimates - i.e., we assumed that parameters $A$, $B$ or $D$
    were zero when making these variance estimates. The lines show the
    true data-only variance (solid), and the result after including
    the first order theoretical corrections to the variance from the
    covariance matrix contribution (dot-dash), the average variance
    estimated naively from the likelihood (dashed) and from the
    distribution of data values that were also used to calculate the
    covariance matrix (dotted).}
  \label{fig:var_toy}
\end{figure}

In order to test the relative methods for determining errors, we have
created Monte-Carlo simulations for a model matching that of
\citet{dodelson13}. Here we assume that each data vector comprised of
$n_b$ values are independently drawn from a standard Gaussian
distribution (mean=0, variance=1), and that $n_s$ of these data
vectors are used to calculate a covariance matrix. The covariance
matrix is allowed to include ``apparent'' correlations between
different data points, even though the true covariance matrix is
diagonal. From any set of data, the parameter we wish to estimate is
the average $p_\alpha$, which has the expected value $E(p_\alpha)=0$,
and true variance $1/n_b$. The 1-dimensional true Fisher matrix and
its inverse are therefore $F_{\alpha\alpha}=n_b$,
$F^{-1}_{\alpha\alpha}=1/n_b$.

We have created $10^5$ Monte-Carlo {\em runs} for every $n_b$ and
$n_s$ tested, averaging the measurements over all runs to provide our
results. For each run, we created a set of $n_s$ data vectors from
which we calculated the covariance matrix and a set of $n_s$
independent data vectors, which we used to test the fit. All of these
data (both dependent and independent data sets) were fitted using the
estimated covariance matrix, using Eq.~(\ref{eq:like}) to estimate the
likelihood. We therefore performed 2$n_s$ likelihood fits for each
run, finding the mean and variance as described in
Section~\ref{sec:like}. Estimates of the variance derived in different
ways from these fits are shown in Fig.~\ref{fig:var_toy}. We do not
apply any bias corrections to these data, but instead plot them as if
they had been naively used to estimate the true variance.

The average variance of the distribution of best-fit parameters
recovered from the fits to the independent sets of data are shown by
the open circles in Fig.~\ref{fig:var_toy}, and are well matched to
the formula derived by \citet{dodelson13} (dot-dash line, given by
Eq.~\ref{eq:var_idist}). These data represent the true error that
should be quoted on measurements. The difference between these data
and the solid line shows the extra variance introduced by the noisy
covariance matrix estimate.

If we estimate the variance using the likelihood, or using the
distribution of data also used to estimate the covariance matrix, we
find a biased value. The average variance recovered by integrating
over the likelihood as in Eq.~(\ref{eq:sig_from_like}) is plotted in
Fig.~\ref{fig:var_toy} (solid circles) - the root of these values are
commonly quoted as parameter errors in analyses. As described in
Section~\ref{sec:like}, for parameters that linearly depend on the
data (or in the standard approximation around the likelihood maxima),
the best-fit value around which we measure the variance does not
matter. Thus we recover exactly the same likelihood errors in our
model whether we use the independent data, or the data also used to
estimate the covariance matrix. These estimates are biased and the
offset is well matched by Eq.~(\ref{eq:var_like}), which is indicated
by the dashed line in the plots. The solid triangles show the variance
estimated from the distribution of best-fit values recovered from the
same data set used to calculate the covariance matrix. These points
are well matched to the dotted line, calculated using the formula
given in Section~\ref{sec:dist_same_data}. As can be seen, this
estimate of the variance is biased low, as a consequence of the offset
between the estimated covariance and the inverse covariance matrix as
given by the extra factor in $m_2$ compared with $m_1$.

In this plot, the factor $m_1$ is the ratio between the dashed and
dot-dashed lines, and $m_2$ is the ratio between dotted and dot-dashed
lines. These factors correct these estimates to produce the true
combined error (dot-dash line) including both the standard variance
and the effect of the noisy covariance matrix.

\section{Cosmological measurements with BOSS 2-point statistics} \label{sec:boss}

\subsection{Estimating the covariance matrix from mocks}

We now apply the calculations described above to investigate
cosmological measurements made with the power spectrum and correlation
function from BOSS. \rev{In this work, we focus on the CMASS galaxy
  sample, although our results could also be applied to the LOWZ
  sample.} BOSS \citep{dawson13} is part of the Sloan Digital Sky
Survey-III (SDSS-III; \citealt{eisenstein11}) project, which used the
SDSS telescope \citep{gunn06} to obtain imaging \citep{gunn98} and
spectroscopic \citep{smee13} data, which was then reduced
\citep{bolton12} to provide a sample of galaxy redshifts, with known
mask. We focus on the BAO methodology described in
\citet{anderson12,anderson13,aardwolf13}, and the RSD methodology of
\citet{reid12,samushia13}.

In \citet{anderson12} and \citet{reid12}, we used $600$ PTHaloe mock
catalogues to analyse the BOSS DR9 sample, both to understand the
analysis methodology and to determine covariance matrices for the
2-point measurements. These mock catalogues were created as described
in \citet{manera13}. Briefly, $600$ 2nd-order Lagrangian Perturbation
Theory (2LPT) matter fields were created in boxes of size
$L=2400\mpcoh$, sampled by $1280^3$ dark matter particles. Within
these boxes, haloes were found with a friends-of-friends group finder
\citep{davis85} with appropriate linking length, and their masses were
calibrated by detailed comparisons with N-body simulations. The halos
were populated with mock galaxies using a Halo Occupation Distribution
(HOD; \citealt{peacock00,cooray02,berlind02}) prescription, which was
calibrated to reproduce the clustering measurements on scales between
$30\mpcoh$ and $80\mpcoh$. Mock catalogues were then created by
sampling these boxes to match the geometry and efficiency of the
project. Mock catalogues have also been drawn from these boxes for the
DR10 \citep{ahn13} and DR11 samples used in \citet{aardwolf13} to measure the BAO
positions \citep{manera13}.

In order to create the mocks, we treat the Northern Galactic Cap (NGC)
and Southern Galactic Cap (SGC) components of the survey as being
independent, and sample them separately from the same set of
boxes. \rev{For the DR9 analysis, we could easily sample the North and
  South components of the survey from the 600 boxes without overlap,
  giving 600 NGC mocks and 600 SGC mocks that are independent.}  Given
the volume covered by the DR10 and DR11 BOSS CMASS galaxy samples, we
could not easily sample both parts of the survey from each box without
overlap, meaning that the NGC and SGC mocks drawn from the same box
are not independent.  To construct joint NGC+SGC mocks, we sample the
NGC from one subset of 300 simulations and combine these with samples
of the SGC from the remaining independent simulations.  An equivalent
set of combined mocks can be created by instead sampling the SGC from
the first subset of 300 simulations and the NGC from the remaining 300
simulations.  While both of these sets should provide unbiased
estimates of the covariance matrix, they are in principle correlated
with each other, \rev{as the set of NGC mocks used to calculate one is
  correlated with the set of SGC mocks used to calculate the other.}
We then estimate the covariance matrix for the \rev{joint NGC+SGC}
power spectrum as the average from these two, \rev{each calculated
  from 300 (NGC+SGC) mock power spectra. The final equation for our
  covariacne matrix is}
\begin{eqnarray} 
  2\hat{C}_{ij} 
  &=&
  \frac{1}{299}\sum_{m<300}[P^m_i(k)-\bar{P_i}(k)][P^m_j(k)-\bar{P_j}(k)]
  \nonumber\\
  &+& 
  \frac{1}{299}\sum_{m>300}[P^m_i(k)-\bar{P_i}(k)][P^m_j(k)-\bar{P_j}(k)],
  \label{eq:C_pk}
\end{eqnarray}
where $P^m_i(k)$ is the measured power spectrum from mock $m$ in bin
$i$, and $\bar{P}_i(k)$ is the mean calculated separately for each set
of 300 mocks. A similar equation is used to calculate the covariance
matrix for the correlation function. Although this approach produces
an unbiased estimate of the covariance matrix, the two contributions
are correlated, so the sum would not produce the $\sqrt{2}$ reduction
in noise in the covariance matrix as would be expected for the
combination of independent estimates, if we could approximate
components as being Gaussian.

\begin{table}
\begin{center}
\begin{tabular}{cccccc}
\hline
sample & \multicolumn{3}{c}{area (deg$^2$)} & $r$ & $(1+r^2)/2$\\
& NGC & SGC & overlap & &\\
\hline
DR9   & 2584 & 690 & 28 & 0.016 & 0.50 \\
DR10 & 4817 & 1345 & 1006 & 0.33 & 0.55 \\
DR11 & 6308 & 2069 & 2069 & 0.49 & 0.62 \\
\hline
\end{tabular}
\end{center}
\caption{The effective areas of the DR9, DR10 and DR11 BOSS CMASS galaxy
  samples, and the overlap areas when mocks are sampled from the same
  parent box. $r$ is the correlation coefficient between estimators and 
  $(1+r^2)/2$ reflects the reduction of the covariance errors
  when the estimators are combined. }
\label{tab:areas}
\end{table}

In fact, for DR9, when projected into the mock boxes, the NGC and SGC
components of the survey only have a small overlap and we were
therefore justified in treating both sets of mocks as
independent. However, for DR10 the overlap is approximately 75 per
cent of the area covered by the SGC, while for DR11 the entire
Southern component is also covered by the NGC (see
Table~\ref{tab:areas}).  If we assume that the variance on the
measurement is proportional to the inverse of the effective volume
$V_{\rm eff}$, the correlation coefficient between \rev{(NGC+SGC)}
mock measurements \rev{with overlapping NGC and SGC components, so
  that the NGC for one overlaps with the SGC of the other, and
  vice-versa,} is given by $r = 2V_{\rm overlap}/(V_{\rm NGC} + V_{\rm
  SGC})$.  The degree of overlap above results in $r=0.33$ for DR10
and $r=0.49$ for DR11.  Again, to be explicit, this correlation
coefficient represents how strongly the power spectrum error in one
mock correlates with that in another mock, where the two mocks sample
either NCG+SCG or SGC+NGC from two simulations.

Ultimately we are interested in how these correlations impact the
covariance error resulting from the combined estimator of
Eq.~(\ref{eq:C_pk}), compared to the covariance error which arises
from using a single set of $300$ mock catalogues.  We find that the
power spectrum correlation propagates into the combined covariance
error, which we effectively model by rescaling the error terms given
by $A$, $B$ and $D$ in Eqns.~(\ref{eq:AB}) \&~(\ref{eq:Psi}) by a
factor of $(1+r^2)/2$\rev{, which is that standard formula for the
  variance of the average of two correlated random variables}.  In the
limit of large $n_s$, the $B$ term dominates over the $A$ term in
Eq.~(\ref{eq:AB}); this is equivalent to rescaling the number of
simulations by a factor of $2/(1+r^2)$.  Thus, for DR9, where the
correlations are negligible, the effect is simply to increase the
effective number of simulations by a factor of two, as one might
expect.

\subsection{Application to DR9 BAO measurements}

\begin{figure*}
    \centering
    \resizebox{0.45\textwidth}{!}{\includegraphics{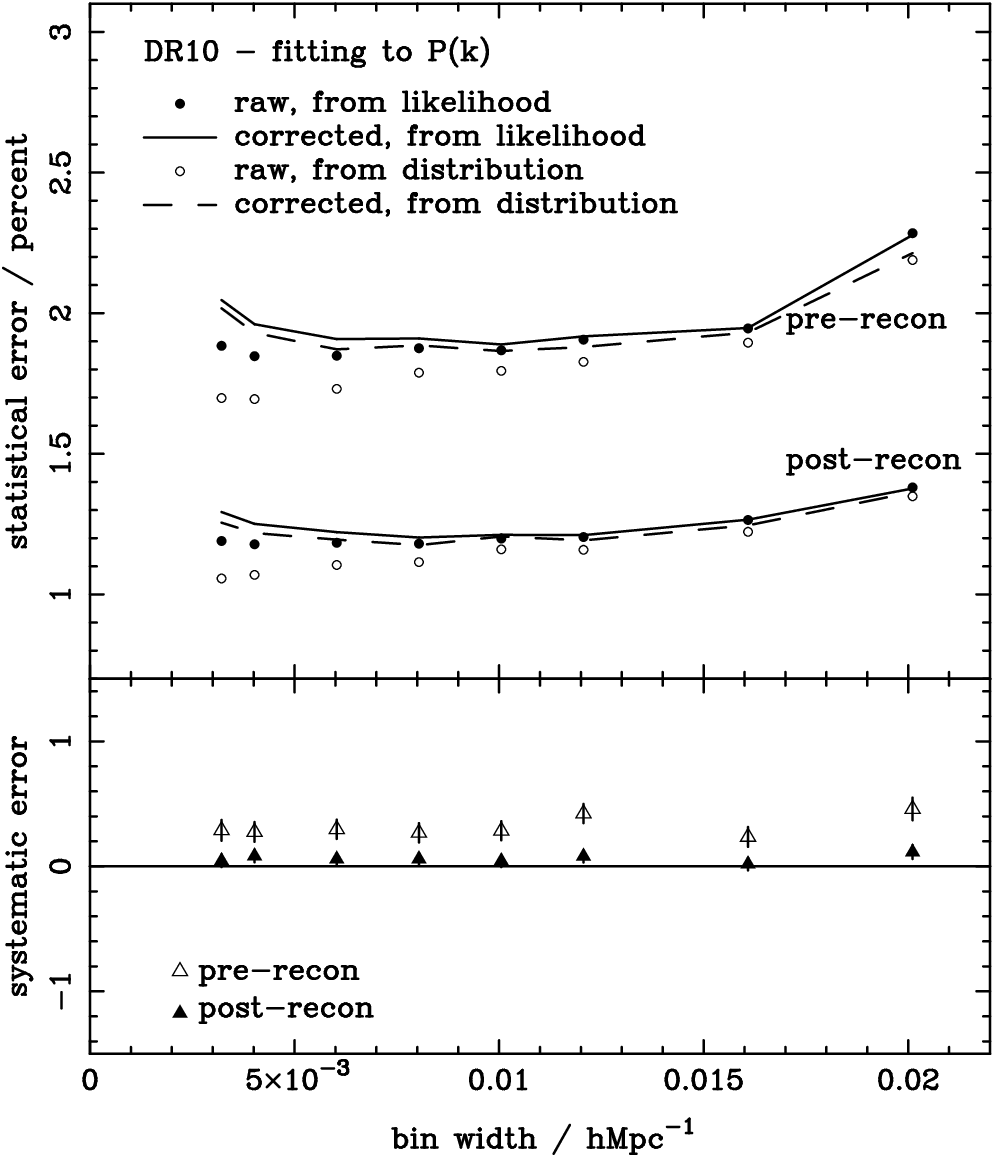}}
    \hspace{1cm}
    \resizebox{0.45\textwidth}{!}{\includegraphics{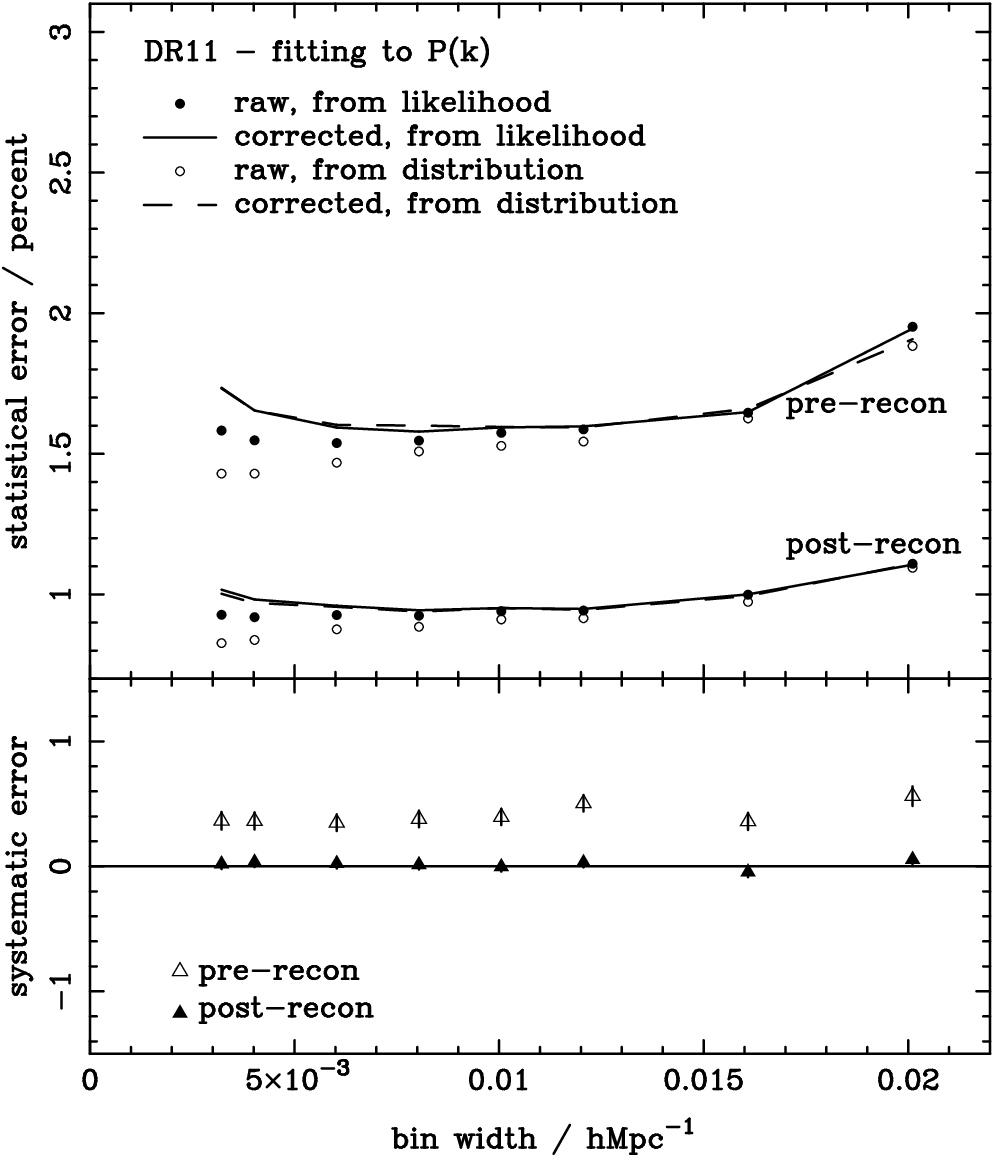}}
    \caption{Top panels: Recovered errors from the best-fit values of
      $\alpha$ calculated by fitting the BAO as described in
      \citet{anderson12}, but for the BOSS DR10 (left) and DR11
      (right) mock samples \citep{manera13b}. Solid circles and the
      solid line were determined from the likelihood, as described in
      Section~\ref{sec:like}, while open circles and the dashed line
      were calculated from the distribution of values recovered from
      the mocks as described in Section~\ref{sec:dist_same_data}. The
      points represent the ``raw'', uncorrected values, while the
      lines show the values after correcting for the covariance
      matrix. Lower panel: Percentage error on the mean value of
      $\alpha$ recovered from the mocks.}
  \label{fig:pk_results}
\end{figure*}

Error-bars for the BAO measurements presented in \citet{anderson12}
were derived from the likelihood calculated from fitting either the
isotropically averaged power spectrum or correlation function with a
model that marginalises out the broadband components of the 2-point
functions, leaving the BAO whose scale can be measured. For the power
spectrum analysis of \citet{anderson12}, we fitted 70 band-powers with
a model including 11 parameters, and neither the correction shown in
Eq.~(\ref{eq:Psi}) nor the factor in Eq.~(\ref{eq:m_1}) were applied
to the inverse covariance matrix. Both the factors are of the order 10
per cent, and act to increase the size of the variance from the raw
value measured. The quoted errors on the power-spectrum based BAO
position measurements provided in \citet{anderson12} should therefore
be increased by 12 per cent given the current analysis: i.e. post
reconstruction, we quoted $\alpha=1.042\pm0.016$, but these will
change with the current error analysis to $\alpha=1.042\pm0.018$.

For the correlation function analysis of \citet{anderson12}, we fitted
44 binned points $28<r<200\mpcoh$, using a model with five free
parameters. As with the fits based on the power spectrum, error-bars
were derived from the likelihood, and neither the correction to the
inverse covariance matrix estimate (Eq.~\ref{eq:Psi}) nor the
correction because of the error in the covariance matrix
(Eq.~\ref{eq:m_1}) were applied. Because of the reduced number of bins
and degrees of freedom, the corrections are slightly smaller than in
the power spectrum case, and are of order 4 per cent and 3 per cent
respectively for the error. The errors on the correlation-function
based BAO position measurements provided in \citet{anderson12} would
therefore need to be increased by 7 per cent given the current
analysis: i.e., post reconstruction, \citet{anderson12} quoted
$\alpha=1.024\pm0.016$, but these values will change with the current
error analysis to $\alpha=1.024\pm0.017$.

\subsection{Application to DR10 and DR11 monopole power spectrum BAO
  measurements} \label{sec:bao_pk}

The default BOSS DR10 analysis presented in \citet{aardwolf13} uses
600 mocks, calculated as for DR9, but with an updated angular mask. We
have measured the power spectrum for each of these mocks after
reconstruction, using the standard pipeline described in
\citet{anderson12}. Each power spectrum was binned into a large number
of fine bins, which were then combined to produce results for various
numbers of bins within the range of scales fitted
$0.02<k<0.3\mpcoh$. For each binning choice, we have estimated the
covariance matrix, and window function, and used these to fit the data
with a model given by
\begin{equation}  \label{eq:mod_pk}
  P^{\rm fit}(k)=P^{\rm sm}(k)
  \left[1+(O^{\rm lin}(k/\alpha)-1)e^{-\frac{1}{2}k^2
      \Sigma_{nl}^2}\right],
\end{equation}
where the BAO scale $\alpha$, and the damping $\Sigma_{nl}$ are
parameters, and $P^{sm}(k)$ is a smooth model for the broad-band shape
of the power spectrum, and $O^{\rm lin}(k)$ are the Baryon
Acoustic Oscillations extracted from the linear power spectrum $P^{\rm
  lin}(k)=O^{\rm lin}(k)P_{\rm sm,lin}(k)$. We have changed the
fitting method from that in \citet{anderson12} in two key ways:
\begin{enumerate}
\item we fit band-powers in $\log P(k)$, which was shown to be close
  to having a multi-variate Gaussian distribution in \citet{ross13a},
  as expected in the sample-variance limited regime.
\item we use a model for the broad band power spectrum 
  \begin{equation}
    P^{\rm sm}(k)= B_p^2P(k)^{\rm sm,lin}+A_1k+A_2+\frac{A_3}{k}+\frac{A_4}{k^2}+\frac{A_5}{k^3},
  \end{equation} 
  which is better matched to that used for the correlation function,
  than the $P(k)$ model used in \citet{anderson12}. Our final model
  has six ``nuisance'' parameters, $B_p$, $A_1$, $A_2$, $A_3$, $A_4$,
  and $A_5$; see \citealt{aardwolf13,ross13b} for further discussion
  of this issue.
\end{enumerate}

For each mock, we have determined the best fit value of $\alpha$ and
$\sigma_\alpha^2$ by marginalising over the other parameters using the
derived likelihood. In this calculation we assumed a Gaussian prior on
$\Sigma_{nl}$ of $\pm2$ centred on the best-fit values determined by
fitting the average recovered power spectrum. \rev{In principle, the
  BOSS data alone can measure this parameter simultaneously with the
  BAO scale measurement, albeit at the expense of an increase in the
  error. In fact we have a strong prior from theory about the
  amplitude of this damping, which we include to reduce the impact on
  the BAO scale error (for more details see \citealt{aardwolf13}).}

We need to apply the corrections determined in Section~\ref{sec:like}
to the errors derived from the likelihood and, as the covariance
matrix was also calculated from the same mocks used to determine the
covariance matrix, the correction of Section~\ref{sec:dist_same_data}
to the distribution of best-fit values. The resulting measurements of
the expected error on $\alpha$ are shown in Fig.~\ref{fig:pk_results}
as a function of the number of bins in the fitted range
$0.02<k<0.3\hompc$. The upper sets of points and lines are
pre-reconstruction, with the lower set, corresponding to the more
accurate fits, post-reconstruction.

The raw errors from these calculations are shown as the open (from the
likelihood) and solid (from the distribution) circles, with the
results after correction represented by the lines. The lower panel
presents the percentage deviation of the mean, calculated from all of the
mocks for different numbers of bins. This represents a systematic
error on the recovered value of $\alpha$.

\begin{figure*}
    \centering
    \resizebox{0.45\textwidth}{!}{\includegraphics{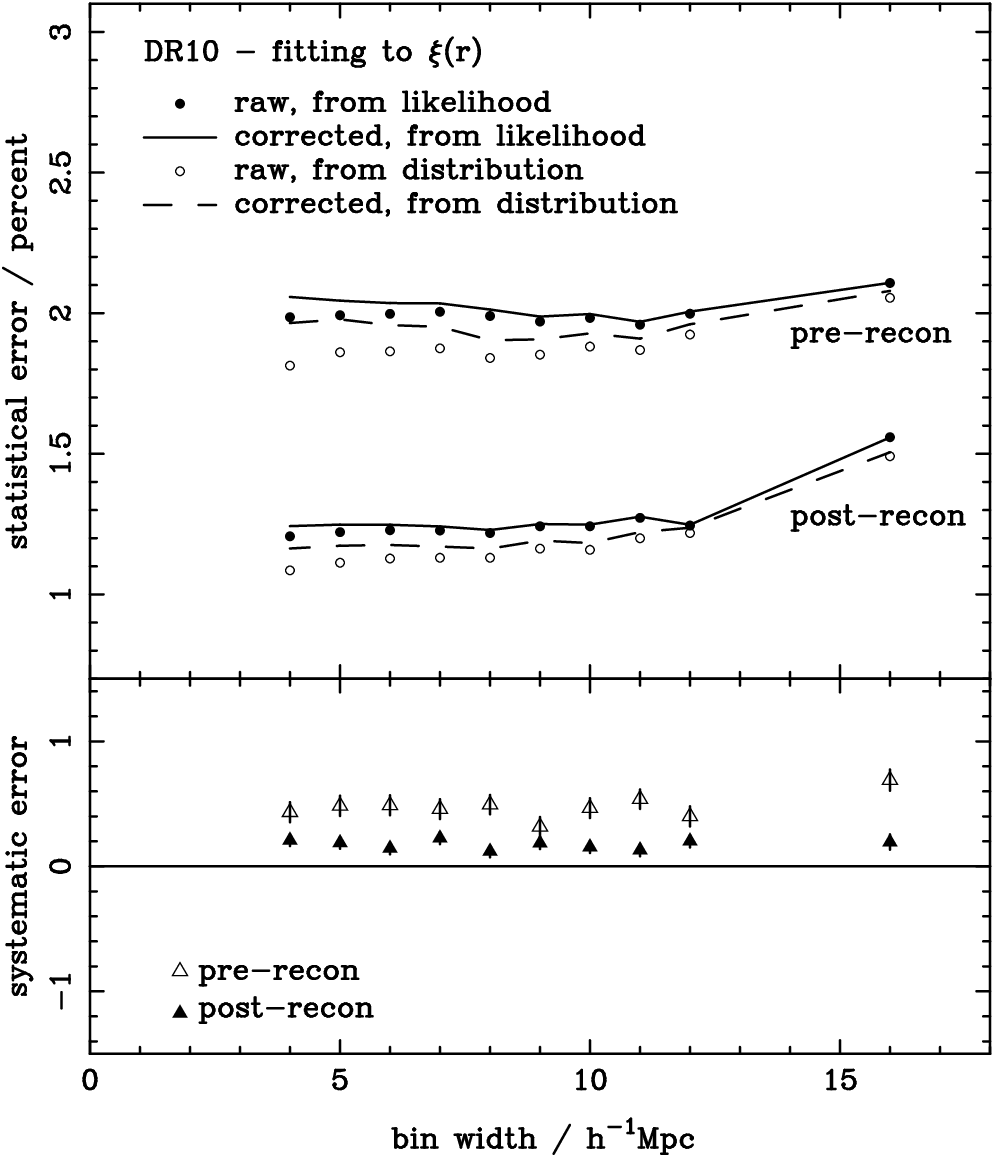}}
    \hspace{1cm}
    \resizebox{0.45\textwidth}{!}{\includegraphics{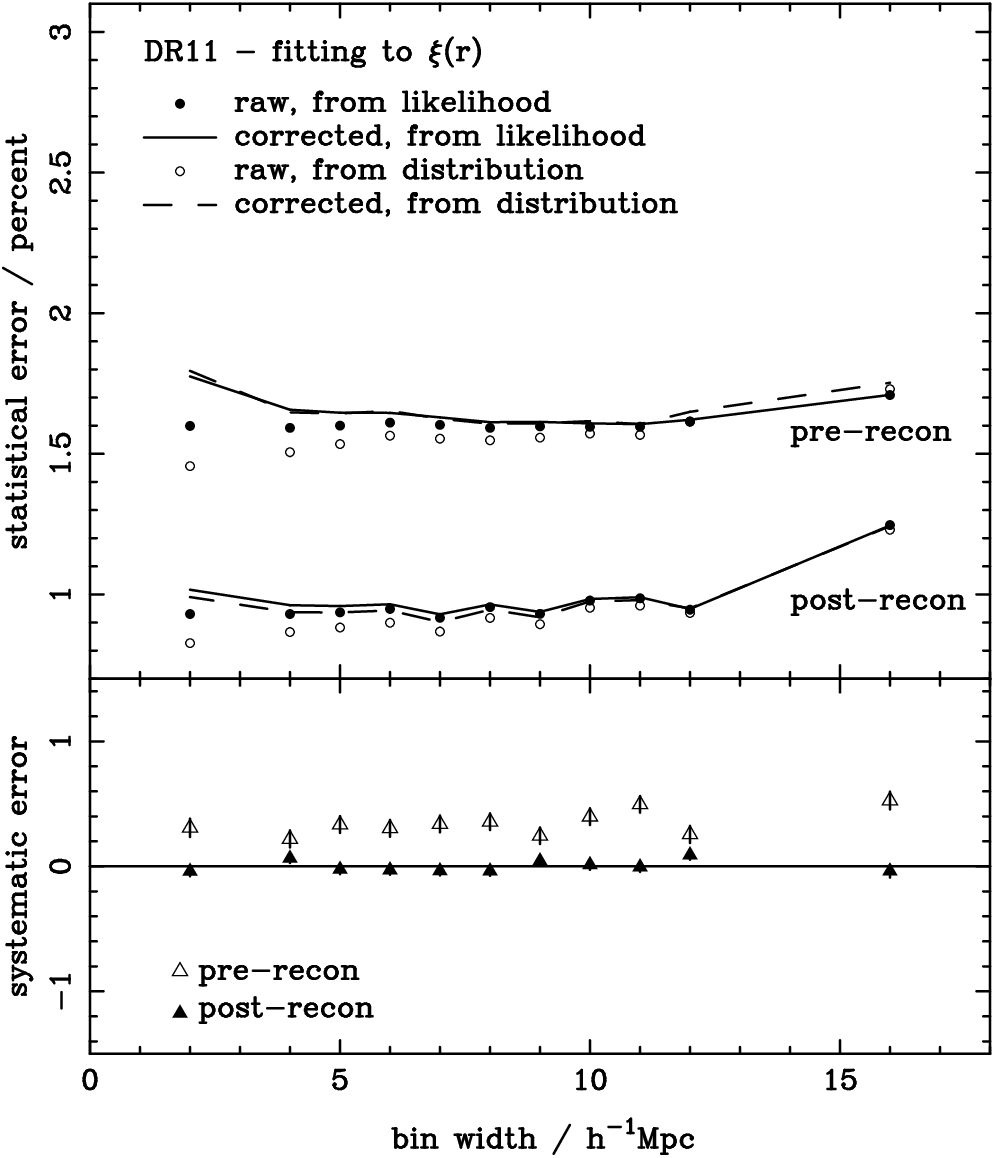}}
    \caption{As Fig.~\ref{fig:pk_results}, but now for the fits to the
      correlation function.}
  \label{fig:xi_results}
\end{figure*}

From Fig.~\ref{fig:pk_results} we see that, after correction, the
values of the errors recovered from the distribution and from the
likelihood agree to a higher degree than before correction,
particularly for small values of the bin width. There is an error on
this match that results from the error in the covariance matrix, with
the data from different bin widths being highly correlated. We expect
this error to be of the same order as the difference between the
corrections applied to the likelihood and distribution based errors as
this difference results from the offset within the Wishart
distribution from which the covariance matrix is derived (see
Section~\ref{sec:intro}). It therefore gives a crude estimate for the
width of this distribution. This reasoning shows that the differences
between corrected errors derived in the different ways are consistent.

Without correction, the statistical errors recovered from both methods
decrease with increasing bin width, naively suggesting that increasing
the number of bins increases the information content. In fact, the
post-correction errors increase for small bin widths, demonstrating
that we are simply transferring data noise into covariance matrix
noise as we increase the number of bins, which is not appearing in the
raw error calculation. \rev{After correction the recovered errors are
  reassuringly independent of bin width for a wide range of bin
  widths.} For small numbers of bins, the mean offset in the BAO
location measured is small compared with the statistical errors, and
is of order 0.4\% for the pre-reconstruction fits, while it is
consistent with zero post-reconstruction, with an error of 0.04\% for
all bin widths. The size of the systematic offset is not dependent on
the bin width, giving us confidence that we are correctly modelling
the binning effects. \rev{The low amplitude of the systematic errors
post-reconstruction} strongly suggests that we do not have any
systematic biases due to the survey mask, our modelling of the
resulting window function, or effects from the galaxy bias as
implemented within the PTHaloe methodology \citep{manera13}.

Comparing both the offset in the mean value recovered and the
recovered errors indicates that the optimum number of bins for the
power spectrum analysis over $0.02<k<0.3\hompc$ is approximately $35$
with bin width $0.008\hompc$, half the number of bins used in the DR9
analysis of \citet{anderson12}. For such a small number of bins, the
corrections required for the derived errors are small:
Eq.~(\ref{eq:m_1}) suggests that the likelihood derived errors need to
increase by $\sqrt{m_1}\sim3$ per cent.

\subsection{Application to DR10 and DR11 monopole correlation function
  BAO measurements} \label{sec:bao_xi}

We have performed a similar analysis to determine the optimum bin size
for the BAO fits to the isotropic correlation function. Following the
methodology adopted for \citet{aardwolf13}, we fix the BAO damping
scale, leaving a 5 parameter model composed of $\alpha$ and a
4-parameter broad-band model that is similar to that described for the
power spectrum in the previous section
\begin{equation}
  \xi^{\rm fit}(s)=B_\xi^2\xi^{\rm mod}(\alpha s)+\frac{a_1}{s^2}+\frac{a_2}{s}+a_3.
\end{equation}
Here, $\xi^{\rm mod}$ is the Fourier transform of a linear model for
the correlation function with damped BAO (see \citealt{aardwolf13} for
more details), and $a_i$ with ($1<i<3$) are free parameters that
marginalise over the broadband signal.  Because the BAO signal in the
correlation function does not extend to non-linear scales to the same
extent as in the power spectrum, the broad-band model can be added to
the linear correlation function, which includes the BAO signal, rather
than multiplying the BAO as in the $P(k)$ model
(Eq.~\ref{eq:mod_pk}). This leaves a correlation-function model with
more freedom to dampen the BAO. Thus, for our correlation function
fits, we fix $\Sigma_{nl}^2$ rather than including it as a free
parameter with a Gaussian prior as for the power spectrum. The
consequences of this difference are discussed further in
\citet{aardwolf13}.

As for the power spectrum, we have fitted all 600 mocks using this
model to determine a likelihood distribution for each. From this
exercise we have derived best-fit values and expected errors on
$\alpha$, marginalising over other parameters. We have also estimated
the width of the distribution of recovered best-fit values, taking
care to produce an unbiased estimate by splitting the mocks into two
sets of 300 independent measurements. The resulting errors, plotted as
a function of number of bins, are shown in
Fig.~\ref{fig:xi_results}. The difference between results from the
likelihood and the distribution are similar to those for the power
spectrum fits. The size of these discrepancies are similar to the
correction applied, and as such may simply be a statistical deviation
within the expected distribution. The results from different bin
choices are obviously correlated to a high degree. We do not attempt
to estimate the error on the correction we are applying to the error -
i.e., the error on the error on the error.

Fig.~\ref{fig:xi_results} reveals a flat minimum, with bins of width
$6$--$10\mpcoh$ all providing similar final errors on the BAO
scale. We therefore recommend that the monopole of the correlation
function, when fitted independently, be binned with width
$8\mpcoh$. There is no evidence that binning on these scales induces a
systematic error due to the coarseness of the averaging.

\begin{figure*}
    \centering
     \resizebox{0.45\textwidth}{!}{\includegraphics{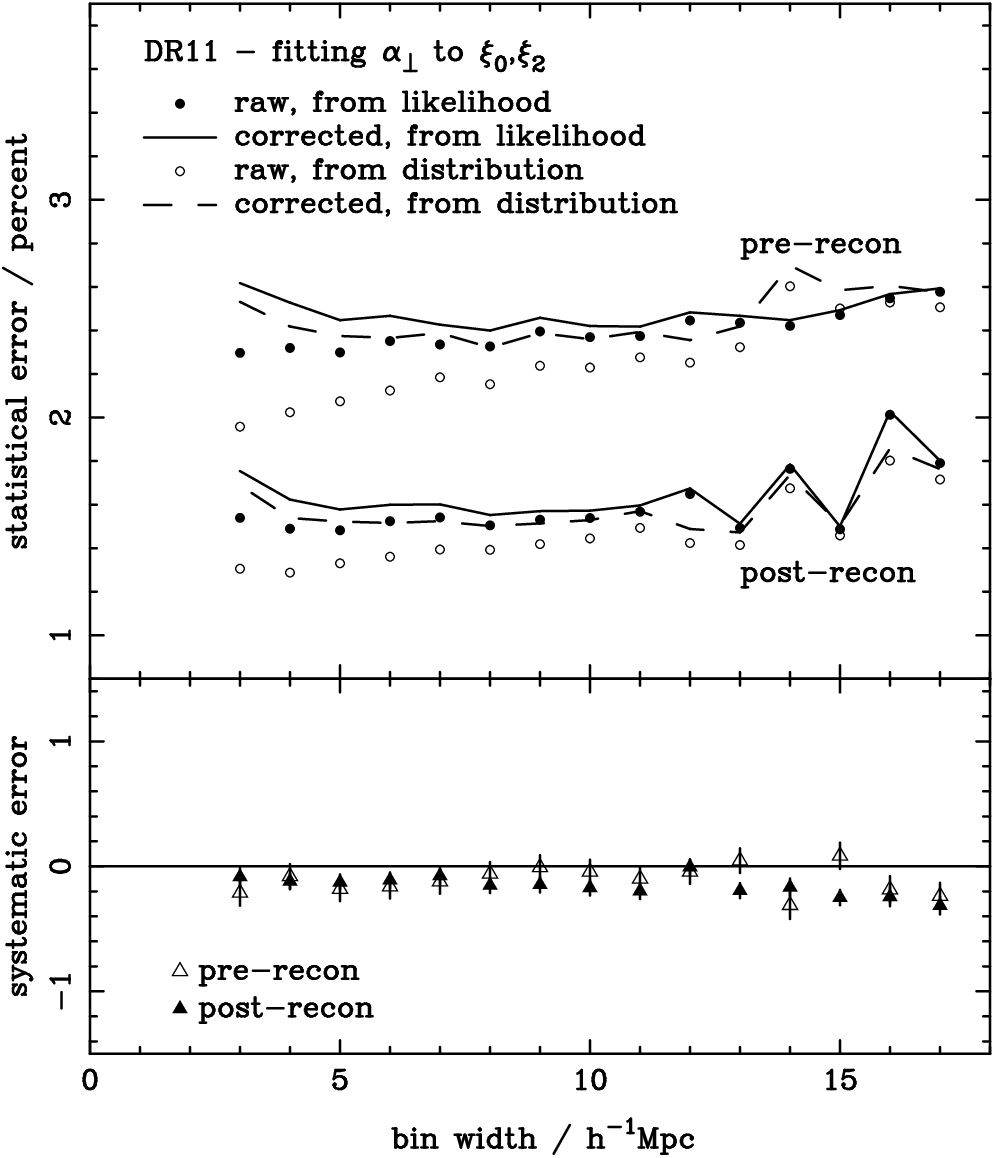}}
    \hspace{1cm}
    \resizebox{0.45\textwidth}{!}{\includegraphics{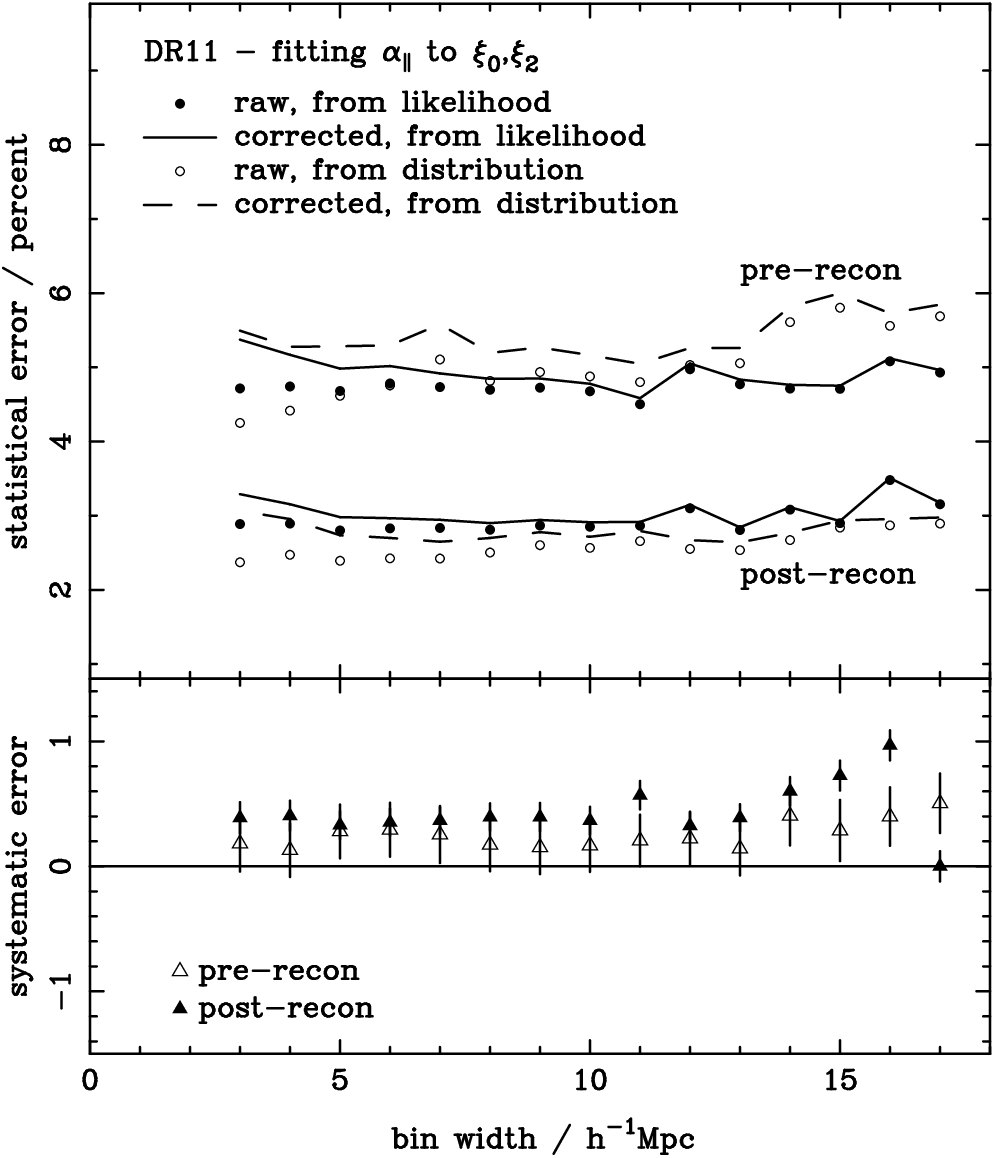}}
    \caption{As Fig.~\ref{fig:pk_results}, but now for BAO fits to
      monopole and quadrupole moments of the correlation function as
      described in \citet{aardwolf13}, now allowing for a different
      dilation of scale in the radial ($\alpha_\parallel$) and angular
      ($\alpha_\perp$) directions.}
  \label{fig:bao2d_results}
\end{figure*}

\subsection{Application to DR11 anisotropic BAO
  measurements} \label{sec:bao2d}

We have also considered fits to the monopole and quadrupole moments of
the correlation function using the methodology applied in
\citet{anderson13} and \citet{aardwolf13}. For simplicity we only
present results from the DR11 data, although similar results are
produced for DR10. Additionally, similar results are observed for fits
to ``Wedges'': top-hat averages of the anisotropic correlation
function in the cosine of the angle to the line-of-sight \citep[for
more information see][]{kazin13}.

Fig.~\ref{fig:bao2d_results} presents the average errors on
$\alpha_\perp$ and $\alpha_\parallel$ from the fits to the 600 mocks
as a function of bin size. As in Fig.~\ref{fig:pk_results}, these
errors are shown with and without the correction factors for the error
in the covariance matrix. The behaviour of the fits in the anisotropic
case is quite similar to those from just fitting the monopole of the
correlation function (Fig.~\ref{fig:xi_results}). The minimum is quite
broad, just pushing to slightly larger bin sizes than the
monopole-only fits. Given our preference for simplicity, we adopt a
bin size of $8\mpcoh$ for fits to both monopole only, or monopole and
quadrupole, rather than using a different bin for the two
measurements.

The likelihood-based and distribution-based results are well matched
after correcting for the covariance matrix effects, as for the
monopole only fits. There is some evidence for a small $\sim$0.5\%
systematic offset on $\alpha_\parallel$, which was also seen in
\citet{anderson13}. \rev{There is also evidence for ``oscillatory
  behaviour'' of the errors as a function of bin width, which is
  particularly apparent for the post-reconstruction fits. For our
  binning scheme, as we increase the bin width, we also alter the
  positions of the bin centres. The ability to fit the position of the
  BAO is very sensitive to the bin centre for bins that cover the BAO
  signal, and are large compared to that signal. This leads to
  variations in the recovered errors as seen. We also see an increase
  in the systematic offset for large bins, which is coupled to this
  lack of resolution. Clearly it is desirable that this region is
  avoided.}

\subsection{Application to DR11 RSD measurements} \label{sec:rsd}

We now extend the analysis to consider RSD measurements made from
joint fits to the monopole and quadrupole moments of the correlation
function. We limit the analysis to have the same bin width for both,
and consider how this choice affects the error on the final
measurement. For this analysis, we have three free parameters: 
\begin{enumerate}
\item The amplitude of the real-space galaxy power spectrum,
  quantified by $b(0.57)\sigma_8(0.57)$, where $\sigma_8(z)$ is the
  root-mean-square amplitude of overdensity fluctuations in spheres of
  radius $8\mpcoh$.
\item The amplitude of the velocity field, which controls the RSD
  amplitude, and is quantified by
\begin{equation}
  f(0.57)\sigma_8(0.57)=\sigma_8(0)\left.\frac{dG}{d\ln
      a}\right|_{z=0.57},
\end{equation}
where $G(z)$ is the linear growth rate
\item The width of the Gaussian probability distribution function
  assumed to model the non-linear Fingers-of-God, $\sigma_{\rm FOG}$.
\end{enumerate}
Further details about these parameters can be found in
\citet{samushia13}. For speed, given the number of fits to be
performed, unlike in \citet{samushia13}, we do not allow the shape of
the real-space power spectrum or the two dilation parameters
$\alpha_\parallel$ and $\alpha_\perp$ that control the radial and
angular projections to vary, and fix them at their true values. We do
not expect this decision to alter our conclusions significantly given
that this shape is highly constrained by the recent Planck results
\citep{planck13}.

\begin{figure}
    \centering
    \resizebox{0.45\textwidth}{!}{\includegraphics{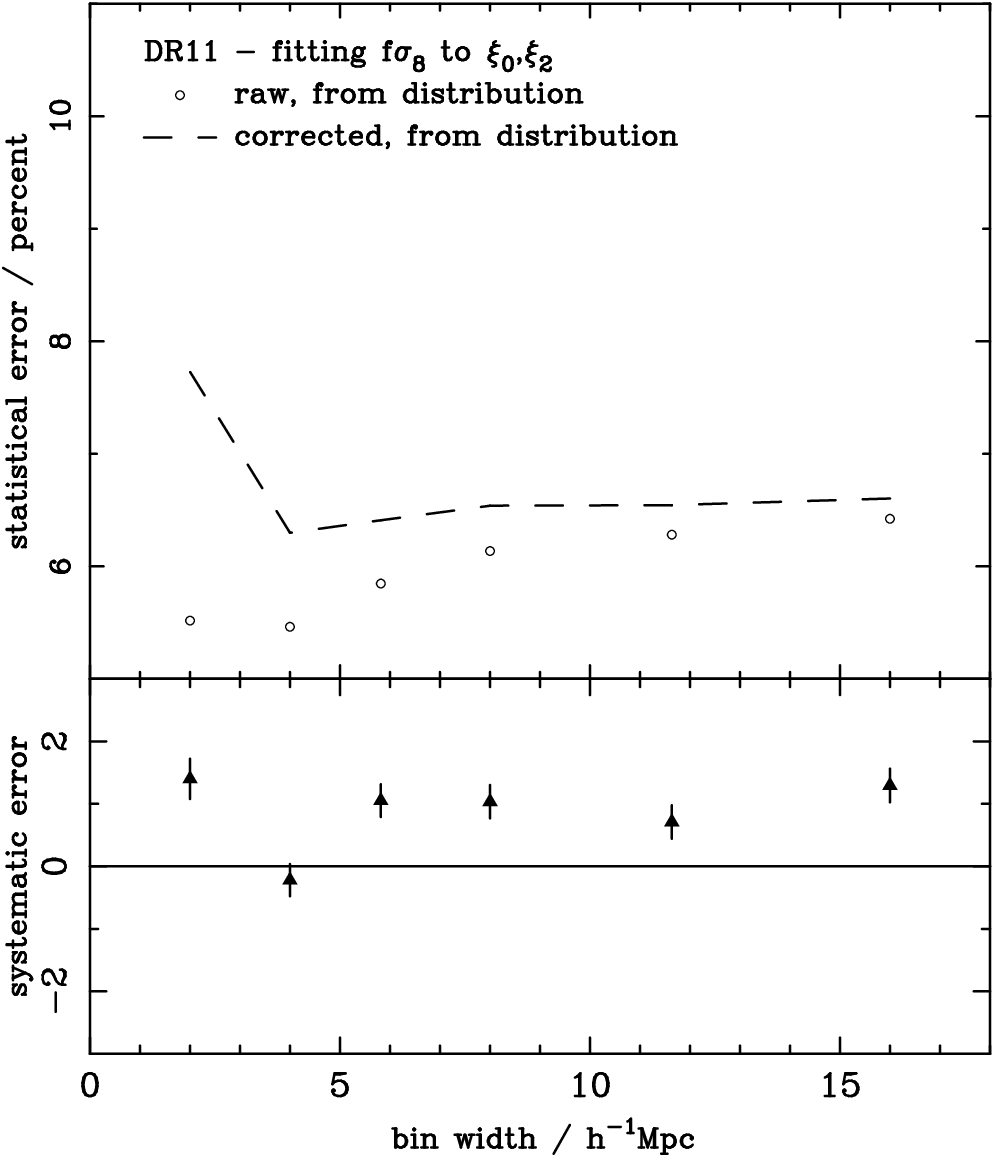}}
    \caption{As Fig.~\ref{fig:pk_results}, but now for RSD fits to
      monopole and quadrupole moments of the correlation function as
      described in \citet{samushia13}.}
  \label{fig:rsd_results}
\end{figure}

The results of our fits can be seen in Fig.~\ref{fig:rsd_results},
where we compare the standard deviations of the distribution of
recovered values of $f\sigma_8$ against bin width. For each fit, we do
not attempt to map the full likelihood, but instead use a minimisation
routine to find the maximum of the likelihood in parameter space. Thus
we only present results from the distribution of recovered best-fit
values. Given the similarity between results derived from individual
likelihood distributions, and from the distributions presented in
Sections~\ref{sec:bao_pk} \&~\ref{sec:bao_xi}, we believe that this
approach is sufficient to determine the best bin width.

As can be seen in Fig.~\ref{fig:rsd_results}, for narrow bin width
where large numbers of bins are used in the covariance matrix, there
is an increase in the corrected error as for the BAO fits. There is no
increase to large bin widths because the RSD measurement is
effectively an amplitude determination unlike BAO fitting, which is a
centroiding problem and therefore large bin widths are more
detrimental. Thus RSD measurements are less sensitive to the bin width
chosen. Most RSD determinations \citep[e.g.,][]{reid12} perform a
joint fit including the shape of the 2-point measurement, and
therefore the best-fit BAO bin width of $\sim8\mpcoh$ remains an
optimal choice. The systematic errors shown in the lower panel of
Fig.~\ref{fig:rsd_results} are relatively large compared with those
from BAO measurements. The large errors are partially due to the 2LPT
mocks not reproducing the nonlinear evolution of the growth rate
exactly. The systematic offset would decrease if we fitted a 2LPT
model to the measurements instead of the nonlinear streaming model,
which is more accurate for the data (see \citealt{samushia13} for more
details).

\section{Discussion}

In this paper we have reviewed the calculations being performed using
the latest BOSS data in order to extract cosmological
measurements. Building upon a series of recent papers examining the
errors in the inverse covariance matrix used in cosmological
applications \citep{hartlap07,taylor12,dodelson13}, we have had to
derive and understand the effect of two further errors in two further
situations - where the error on final parameters is calculated by
integrating over the derived likelihood and, in order to test the
method, the distribution of best-fit values recovered from the same
set of mocks used to determine the covariance matrix. These
derivations have been tested, and shown to be accurate using
Monte-Carlo simulations.

To summarise, there are two corrections that must be applied to the
``naive'' analysis simply inverting the covariance matrix derived from
Eq.~(\ref{eq:cov_estimate}), and using it in Eqns.~(\ref{eq:like})
\&~(\ref{eq:chisq}). First, as pointed out by \citet{hartlap07}, we
must correct for the offset nature of the Inverse Wishart distribution
by correcting the inverse covariance matrix by the factor given in
Eq.~(\ref{eq:Psi}). Second, we need to correct for the additional
contribution of the error in the covariance matrix to the final error
on a derived parameter. Three different corrections to create unbiased
error estimates exist in different situations:
\begin{enumerate}
\item If the variance of a measurement is estimated from the
  distribution of best-fit values recovered from data that are
  independent of that used to estimate the covariance matrix, the
  variance on the result is given in Eq.~(\ref{eq:var_idist})
  \citep{dodelson13}. This variance corresponds to the true error on
  measurement from data (which are independent from the mocks used to
  calculate the covariance matrix).
\item If the variance is measured from a likelihood, calculated from
  fitting to a set of data (be it from independent mocks, the same
  mocks used to estimate the covariance matrix, or the actual data),
  we derive a biased estimate of the variance, which is different from
  the expression given by Eq.~(\ref{eq:var_idist}). To correct back to
  this variance, we must apply the correction $m_1$, given in
  Eq.~(\ref{eq:m_1}) to the derived estimate.
\item If the variance is derived from the distribution of best-fit
  values recovered from the same data also used to estimate the
  covariance matrix, we also obtain a biased result, and must now
  apply the factor $m_2$ given in Eq.~(\ref{eq:m_2}) to the estimate.
\end{enumerate}

We have considered how the mocks used to determine the covariances for
BOSS affect parameter inferences, and have shown how they must be
carefully analysed in order to take into account how they were
produced, in particular the overlap between NGC and SGC
components. Having done this, we have not only included the extra
errors in our final measurement errors given in companion papers
\citep{aardwolf13,Beutler13,Chuang13,samushia13,sanchez13,Tojeiro13},
but also used the derivation to understand the effect of bin size on
the final errors. We have derived optimal binning strategies for BAO
fits to the monopole correlation function and isotropically-averaged
power spectrum, and anisotropic BAO fits and RSD fits to the monopole
and quadrupole moments of the correlation function. \rev{These
  best-fit strategies are dependent on the level of precision achieved
  within the covariance matrix. If more mocks were used, or higher
  precision could be achieved in some other way, then fits using more
  bins would become more desirable. However, after applying all
  corrections, the isotropically averaged BAO distance scale error
  recovered from the mocks is quite independent of bin size over a
  broad range of bin widths. This suggests that our best strategy will
  not change significantly even with better precision for the
  covariance matrix. The lack of sensitivity to bin size is good to
  see, as one would hope that the analysis method does not have a
  strong effect on the final measurements. The ability to recover the
  BAO scale without significant loss of accuracy using large bin sizes
  up to $12\mpcoh$ for $\xi(s)$, is perhaps more surprising, although we
  note that the BAO feature is quite broad.}

Our analysis on bin sizes demonstrates that, on average, after
correction, the recovered errors derived in multiple ways are a better
match to each other than before correction. However, we caution that
this match depends on the actual noise in the covariance matrix, which
might be expected to be of the same order as the difference between
correction factors. This match also relies on the model adopted being
a good fit to the data. For the fit to BAO positions it is clear that
a poor model can yield incorrect likelihood errors, while leaving the
distribution of best-fit values relatively unaffected. The damping
term in Eq.~\ref{eq:mod_pk} is critical here - for any fit to data, if
the model is over-damped, the likelihood maximum will be reduced as
the model has more freedom to move, although the best-fit location for
each mock will generally not change by the same amount. For an
under-damped model, the likelihood maximum will be increased, although
the data themselves do not support such an apparent improvement in
errors, as evidenced by the recovered distribution of best-fit
values. Further investigation is required, but is outside of the remit
of this paper.

The comparison of BAO measurement errors as a function of bin size
raises the interesting question of why the corrected error increases
for increasing numbers of bins. The covariance matrix for large
numbers of bins obviously still contains all the information used with
a smaller number of bins, so theoretically you should be able to
extract the same information from it and the data. As discussed in
Section~\ref{sec:intro}, the correct approach is to construct a joint
likelihood of the data and mocks given the cosmological model to be
tested. In the standard Gaussian assumption on the distributions of
mocks and data, this is the same as that given by
Eq.~(\ref{eq:like_full}). Marginalising over the true covariance
matrix would then yield the final likelihood for the parameters given
the mocks - in essence this should be the same for any bin choice for
smooth models, where the binning results in minimal loss of
information. The problem is that we're not performing this optimal
likelihood approach if we assume that the estimated covariance matrix
is "correct" and use the standard likelihood equation
(Eq.~\ref{eq:like}). In this approach, the effect of the covariance
matrix, and the error it introduces through Eq.~(\ref{eq:like}),
changes with bin size: this dependence is given in
Eqns.~(\ref{eq:DPsi}) \&~(\ref{eq:AB}), and is propagated through to
the final error on the recovered parameters. Thus, the optimal bin
size is actually only an optimal bin size if you want to retain
Eq.~(\ref{eq:like}) as the likelihood equation - in this case the
error does depend on bin size, and the error increases with increasing
number of bins.  The increase to large bin sizes can be more easily
understood - here we are simply losing information as the averaging
being performed increases in importance, leading to increasing
errors. The minimum in the recovered error balances these two effects.

In sections~\ref{sec:bao2d} \&~\ref{sec:rsd}, we saw that the
corrections required to the combined fits to both the monopole and
quadrupole are quite large for both BAO and RSD measurements. This
result suggests that there are significant gains to be obtained either
creating more accurate covariance matrices, or by reworking the
likelihood calculation to include covariance matrix errors. For future
surveys, this effect will become increasingly important\rev{, and having
too few mocks, or too poor a model for the covariance matrix will have
a serious impact on the measurements made}.

\section*{Acknowledgements}

WJP acknowledges support from the UK Science \& Technology Facilities
Council (STFC) through the consolidated grant ST/K0090X/1, and from
the European Research Council through the ``Starting Independent
Research'' grant 202686, MDEPUGS. AGS acknowledges support from the
Trans-regional Collaborative Research Centre TR33 `The Dark Universe'
of the German Research Foundation (DFG). Power spectrum calculations
and fits made use of the COSMOS/Universe super-computer, a UK-Dirac
facility supported by HEFCE and STFC in cooperation with CGI/Intel.

Funding for SDSS-III has been provided by the Alfred P. Sloan
Foundation, the Participating Institutions, the National Science
Foundation, and the U.S. Department of Energy Office of Science. The
SDSS-III web site is http://www.sdss3.org/.

SDSS-III is managed by the Astrophysical Research Consortium for the
Participating Institutions of the SDSS-III Collaboration including the
University of Arizona, the Brazilian Participation Group, Brookhaven
National Laboratory, University of Cambridge, Carnegie Mellon
University, University of Florida, the French Participation Group, the
German Participation Group, Harvard University, the Instituto de
Astrofisica de Canarias, the Michigan State/Notre Dame/JINA
Participation Group, Johns Hopkins University, Lawrence Berkeley
National Laboratory, Max Planck Institute for Astrophysics, Max Planck
Institute for Extraterrestrial Physics, New Mexico State University,
New York University, Ohio State University, Pennsylvania State
University, University of Portsmouth, Princeton University, the
Spanish Participation Group, University of Tokyo, University of Utah,
Vanderbilt University, University of Virginia, University of
Washington, and Yale University.

\label{lastpage}

\end{document}